# Breakup of dense colloidal aggregates under hydrodynamic stresses

Alessio Zaccone, Miroslav Soos, Marco Lattuada, Hua Wu, Matthäus U. Bäbler, and

Massimo Morbidelli

Institute for Chemical and Bioengineering,

Department of Chemistry and Applied Biosciences

ETH Zurich, 8093 Zurich, Switzerland

CORRESPONDING AUTHOR

Massimo Morbidelli

Email: massimo.morbidelli@chem.ethz.ch.

Fax: 0041-44-6321082.

**ABSTRACT**

Flow-induced aggregation of colloidal particles leads to aggregates with fairly high fractal dimension ($d_f \simeq 2.4-3.0$) which are directly responsible for the observed rheological properties of sheared dispersions. We address the problem of the decrease of aggregate size with increasing hydrodynamic stress, as a consequence of breakup, by means of a fracture-mechanics model complemented by experiments in a multi-pass extensional (laminar) flow device. Evidence is shown that as long as the inner density decay with linear size within the aggregate (due to fractality) is not negligible (as for $d_f \simeq 2.4-2.8$), this imposes a substantial limitation to the hydrodynamic fragmentation process as compared with non-fractal aggregates (where the critical stress is practically size-independent). This is due to the fact that breaking up a fractal object leads to denser fractals which better withstand stress. In turbulent flows, accounting for intermittency introduces just a small deviation with respect to the laminar case, while the model predictions are equally in good agreement with experiments from the literature. Our findings are summarized in a diagram for the breakup exponent (governing the size versus stress scaling) as a function of fractal dimension.

# I. INTRODUCTION

In phase transitions dynamics, the formation of either stable phases or states under (quasi)equilibrium conditions, as well as the formation of metastable states, has been the object of intense study in the past. However, in many application as well as natural contexts, nucleation and growth occur under *driven* conditions, for instance under an imposed field of shear [1]. As shown in recent work [2], even in the simplest systems the presence of shear leads to complex nucleation behaviours by affecting the rate of nucleation, the growth of the aggregates, and their breakup. Here, it is our aim to rationalize the effect of hydrodynamic stresses, in either laminar or turbulent regime, on the (mechanical) stability and breakup of dense colloidal aggregates formed under shearing conditions, and the interplay with their fractal or non-fractal morphology. Unlike the case of emulsions, where the stability with respect to flow-induced breakup of single drops can be straightforwardly evaluated (even under turbulent conditions), in terms of a balance between viscous and capillary forces (see [3] and references therein), the situation for colloidal aggregates is more complicated. This is due to the more complex structure of dense colloidal aggregates and especially to their response to stress, which has remained hitherto elusive, despite its crucial role in the rheological behaviour of colloidal gels and glasses, as recent works have suggested [4]. In particular, the balance between clustering (structure-formation) and breakup (structure-failure) is essential to understand the structural origin of the puzzling rheological properties of complex fluids, such as *rheopexy* (the increase of viscosity with time under steady shearing), often found in biological fluids, and its counterpart, *thixotropy* (the decrease of viscosity with shearing time) [4]. Our aim here is to provide a physical description of the hydrodynamic failure of colloidal aggregates without which no microscopic understanding of complex fluid rheology is possible.

## II. MODEL DEVELOPMENT

Thus, we start from basic considerations on the structure of colloidal aggregates formed in flows at high Peclet numbers, where aggregation is entirely dominated by convection while Brownian motion has a negligible effect. Moreover, coagulation in the primary minimum of interaction energy is assumed. Aggregation mechanism is therefore essentially ballistic, i.e. a superposition of ballistic particle-cluster and ballistic cluster-cluster aggregation mechanisms. The first mechanism results in the formation of aggregates with fractal dimension up to $d_f = d = 3$ (i.e. homogeneous) in 3D, while the second mechanism yields aggregates with $d_f \simeq 2$. In reality it has been shown, within numerical studies, that flow-induced aggregation events involve pairs of aggregates which may exhibit significant disparity in size (thus resembling particle-cluster ballistic aggregation) although the resulting fractal dimension, due to the flow streamlines getting screened from the interior of the larger aggregates, is lower and typically $d_f \simeq 2.5 - 2.6$ [5]. These values from numerics [5] agree well with measured values which fall within the range $2.4 < d_f < 2.8$ [6]. Aggregates with such fairly high fractal dimension are dense and almost compact, although not homogeneous, thus quite different from the more studied fractals featuring $d_f \lesssim 2$, for which fractal elastic models, explicitly accounting for the fractal nature of the stress-transmission network, have been proposed [7]. In the following, we will use the term *fractal* essentially to designate the power-law variation of volume fraction in the aggregate as $\phi(r) \sim r^{-(d-d_f)}$, $r$ being the radial coordinate measured from the centre of mass of the aggregate, which defines $d_f$ as what we call the aggregate fractal dimension. Such dense fractals are found in a variety of physical systems, e.g. the compaction of nanometer silica particles [8], the aggregation of proteins [9], the structure of proteins themselves [10], and as the building blocks of metallic glasses, as suggested in [11]. Recently, it has been observed that, in attractive colloids, under certain conditions the interplay between spinodal phase separation and glass transition leads to fractal aggregates with $2.4 < d_f < 2.6$ and inner (average) volume fraction $\phi \simeq 0.5$ [12]. Therefore, such aggregates are internally amorphous (i.e. they possess the short-range structure of liquids and glasses, with no long-range order), at the same time exhibiting power-law decay of density with linear size. Aggregates formed under flow, which possess a fractal dimension in the same range, exhibit a similar amorphous character which is quite evident e.g. in the micrographs of Ref. [6]. Motivated by these considerations and by the recent observation that sufficiently large, dense colloidal aggregates break up in shear flow by unstable propagation of cracks [13], we propose a mean-field-like criterion for breakup in analogy with amorphous solids. Based on energy conservation, the (Griffith-like) critical condition for breakup is given by equating the strain energy supplied by the (external) hydrodynamic stress $d\mathcal{E}_1 \simeq (\sigma^2/2E)\xi^d$, to the energy required to extend the fracture surface $d\mathcal{E}_2 \simeq \Gamma \xi^{d-1}$. Here, $E$ is the Young's modulus of the aggregate, $\sigma$ is the applied stress, $\xi$ is the characteristic radius of the initial crack (typically of cusps on the aggregate surface), and $\Gamma$ the surface energy associated with the broken bonds in extending the crack surface [14]. This yields the following relation for the critical stress

$$\sigma^2 \simeq E\Gamma\xi^{-1} \qquad (1)$$

in analogy with disordered solids [14], where $\sigma$ is the critical value of applied stress. Each term on the r.h.s of Eq. (1) is a function of the average particle volume fraction in the aggregate, $\phi$. In recent work [15], the shear modulus of amorphous solids made of particles interacting via both central and tangential (bond-bending) interactions has been derived systematically using Alexander's Cauchy-Born approach [16] in the continuum limit. The employed affine approximation is expected to yield small errors for overconstrained (hyperstatic) packings. Such situation occurs in covalent glasses (e.g. Ge, Si), where covalent bonds (which can support significant bending moments) are very effective in reducing the number of degrees of freedom per particle. A similar situation is encountered with coagulated colloids [17], especially polymer colloids where mechanical adhesion of the interparticle stabilizes them against tangential sliding [17]. According to [15], the shear modulus for a glass of spherical particles interacting via central (C) as well as bond-bending (B) interactions can be estimated as $G \simeq \left[(4/5\pi)\kappa_{\parallel}z^{(C)} + (124/135\pi)\kappa_{\perp}z^{(B)}\right]\phi R_0^{2-d}$, where $\kappa_{\parallel}$ and $\kappa_{\perp}$ are the bond stiffness coefficients for central and bond-bending interactions respectively, $R_0$ is the distance between bonded particles in the reference configuration (defined as in Cauchy-Born theory [15]), and $z$ is the mean coordination number. In a system of like particles with bond-bending resistance, it is $z = z^{(C)} = z^{(B)}$, and the volume fraction scaling is thus given by

$$E \sim G \sim \phi z(\phi) \qquad (2)$$

In the absence of strong interparticle bonds, affinity is reasonable only for significantly overconstrained packings, i.e. well above the (geometric) rigidity threshold or isostatic point, $J$, where the number of geometric constraints just equals the number of degrees of freedom (the latter given by $2d$). Close to the isostatic transition, the proper (critical) scaling is $G \sim \phi_J[z(\phi) - z_J]$, where $z_J$ and $\phi_J$ are the mean coordination and the volume fraction at point $J$ [15, 18]. Eq. (2) is reasonably valid in the case we are considering here of dense aggregates ($\phi \gtrsim 0.4$) where interparticle bonds can sustain significant bending moments since already with $z^{(B)} = z^{(C)} = 3$ the system is largely overconstrained (the number of saturated degrees of freedom being equal to $z^{(C)}/2 + z^{(B)}(z^{(B)} - 1)/2 = 9$) and nonaffine displacements are small. The evolution with $\phi$ of the mean coordination (within the aggregate) can be estimated in analogy with deeply quenched, dense monoatomic glasses (for at such high density the structure is dominated by the hard-sphere component of interaction). Thus, we integrate the radial distribution function $g(l)$ of hard-sphere liquids, $z(l^\dagger; \phi) \simeq 24\phi \int_0^{l^\dagger} (1+l)^2 g(l; \phi) dl$, with a cut-off $l^\dagger$ determined by the isostatic point of hard-spheres ($\phi \simeq 0.64$). For $g(l; \phi)$ we use standard liquid theory, with the Verlet-Weis correction and the Hall equation of state valid in the dense hard-sphere fluid [19]. In the glassy regime of interest here ($0.5 \lesssim \phi \lesssim 0.6$), the so obtained $z = z(\phi)$ can be approximated with a power-law with good accuracy ($R^2 = 0.993$) yielding $z \sim \phi^\beta$, with $\beta \simeq 3.8$ (see [19] for the full derivation and details), so that

$$E \sim \phi^{\beta+1}, \quad \beta \simeq 3.8 \tag{3}$$

From observations on a similar system (a disordered agglomerate of particles with mechanical adhesion in a dense range of $\phi$ starting from $\phi \simeq 0.49$), Shahidzadeh-Bonn et al. [20] have shown that the surface energy term $\Gamma$ obeys the same dependence on volume fraction as the elastic modulus,

$$\Gamma \sim E \sim \phi^{\beta+1} \tag{4}$$

The same relation may be obtained by Cauchy-Born expanding (in 2D) the free energy of the fracture surface [21]. Finally, the initial size of the crack ($\xi$) is a decreasing function of $\phi$. A precise determination of this dependence is non-trivial. Here, we should content ourselves with two limiting cases. One is the case of a fully-developed fractal object where the simplest meaningful *ansatz* is $\xi \sim L$. This is equivalent to observing that the size of the initial crack be proportional to the characteristic size of the aggregate, *L*. As previously mentioned, the local volume fraction in the aggregate scales with the radial coordinate (*r*) as $\phi(r) \sim r^{-(d-d_f)}$, so that the average volume fraction in the aggregate obeys $\phi \sim L^{-(d-d_f)}$, or, in terms of the aggregate radius of gyration $\phi \sim R_g^{-(d-d_f)}$, leading to $\xi \sim L \sim \phi^{-1/(d-d_f)}$. Use of the latter (fractal) scaling and combination of (2)-(4) into (1), lead to the following power-law scaling for the critical stress required to initiate breakup

$$\sigma \sim R_g^{-(1/2)(d-d_f)[2(\beta+1)+1/(d-d_f)]} \tag{5}$$

In the other limit of homogeneous (non-fractal) solids, the relation between the average crack size and the volume fraction is of direct proportionality. This may be seen if one treats the initial cracks as inclusions of effective size $\xi$, such that $\xi \sim (1-\phi)^3$. It can be easily verified that in the range $0.4 < \phi < 0.7$ the latter relation is practically equivalent to the relation $\xi \sim \phi^{-0.4}$, found within well-known studies of disordered solids in the past [22]. Thus, for aggregates with high fractal dimension close to the limit $d_f \to d$, we expect the relation $\sigma \sim R_g^{-(1/2)(d-d_f)[2(\beta+1)+0.4]}$ to apply, instead of Eq. (5). Generalizing, we will have

$$\sigma \sim R_g^{-(1/2)(d-d_f)[2(\beta+1)-\nu]} \tag{6-a}$$

with either

$$\nu = -1/(d-d_f) \quad \text{or} \quad \nu \simeq -0.4 \tag{6-b}$$

in $\xi \sim \phi^\nu$ depending on whether the cracking is dominated, respectively, by fractality or by quasi-homogeneous structural disorder.

Thus, considering that for spatially heterogeneous flows all aggregates are exposed to the highest flow intensity regions only in the steady-state limit ($t \to \infty$), the following scaling for the steady-state aggregate size ($R_g$) and mass ($X \sim R_g^{d_f}$) that are mechanically stable in heterogeneous flows at a given stress $\sigma$ is finally derived as

$$\lim_{t \to \infty} R_g(t) \equiv R_g^{(s)} \sim \sigma^{-2/(d-d_f)[2(\beta+1)-\nu]} = \sigma^p \tag{7-a}$$

$$\lim_{t \to \infty} X(t) \equiv X^{(s)} \sim \sigma^{-2d_f/(d-d_f)[2(\beta+1)-\nu]} = \sigma^{d_f p}, \tag{7-b}$$

with

$$p = -2/(d-d_f)[2(\beta+1)-\nu]. \tag{7-c}$$

### III. EXPERIMENTS, COMPARISON WITH THE MODEL, AND DISCUSSION

#### A. Laminar flow

To test these predictions we have carried out experiments using colloidal aggregates generated in a stirred vessel with a well-characterized flow field from fully-destabilized polystyrene particles of radius $a \simeq 405$ nm (Interfacial Dynamics, USA), with $d_f = 2.69 \pm 0.2$, from optical microscopy, according to the procedure reported in [23]. Detailed description of the materials, methods, and devices can be found in [24].

The aggregate suspension, under very dilute conditions (total solid fraction of the aggregate dispersion equals $2\times10^{-6}$) to avoid further aggregation during the flow experiment, were subsequently injected into a channel with a restriction in the middle (convergent-divergent nozzle) which allows achieving a substantially intense flow (the highest velocity gradients being near the entrance of the nozzle), as sketched in Fig. 1. The contraction radius, $D_n$, was varied in the range 0.25-1.5 mm. The extensional flow field realised in the nozzle has been thoroughly characterized by numerically solving Navier-Stokes equations using a CFD code whereby it is shown that under all conditions the flow of the restriction entrance is laminar [24]. The resulting contour plots of hydrodynamic stresses are shown elsewhere [24]. Conditions of laminar flow are ensured when $\text{Re} \lesssim 1000$. The steady-state average gyration radius, $R_g^{(s)}$, and average zero-angle intensity of scattered light, $I(0)$, (the average is meant over the population of aggregates) were measured off-site by small angle light scattering (SALS), Mastersizer 2000 (Malvern, U.K.) under very dilute conditions (total volume fraction of the dispersion $\sim 10^{-5}$). By repeatedly passing the aggregate suspension through the nozzle, a stationary condition is achieved where the average aggregate size reaches a steady value, independent of the number of passes. $R_g^{(s)}$ values have been plotted in Fig. 2(a) as a function of the stress acting on the aggregate, $\sigma$, which is identified with the hydrodynamic stress resulting from the mean velocity gradient $\sigma \simeq (5/2)\mu\alpha_L$, where $\alpha_L$ is the highest positive eigenvalue of the velocity gradient tensor (evaluated from CFD calculations [24]), and $\mu$ is the fluid viscosity. The experimental data are thus compared with predictions of Eq. (7)-a using the experimentally determined value $d_f = 2.7$ which does not change appreciably with increasing flow intensity, as found also in previous work [7,24]. The agreement between the scaling predictions of Eq. (7)-a and the experimental trend is strikingly good. However, considering that uncertainty on the experimentally determined fractal dimension is high, we have analyzed the scattering properties of the aggregates and calculated the zero-angle intensity of scattered light (proportional to the aggregate mass, $I(0) \sim X^{(s)}$) of computer-generated aggregates with tuneable fractal dimension (where a Voronoi tessellation-based densification algorithm was employed to generate clusters with $d_f > 2.5$, as described in [25]). Mean-field T-matrix theory [26] has been used to account for multiple-scattering (which is particularly strong with such dense aggregates). Further, for the size of the computer-generated aggregates we used $R_g^{(s)}$ values from fitting Eq. (7)-a to the experimental data. The optimum *quantitative* agreement, shown in Fig. 2(b), between the calculated and the measured $I(0)$ is obtained when computer-generated aggregates with $d_f = 2.7$ are used for the calculation (some offset at large $I(0)$ values cannot be avoided and is ascribed to the size polydispersity of the aggregate population in that regime). This confirms the value measured experimentally and justifies its use for the comparison between model and experiments in Fig. 2(a). Predictions of the $I(0)$ versus stress scaling using Eq. (7)-b, in accord with Rayleigh-Debye-Gans (RDG) theory of light scattering, which neglects multiple-scattering, are shown in Fig. 2(b). They lie very far from the experimental data, thus indicating the importance of multiple-scattering effects in the system.

**B. Turbulent flow**

Let us now consider a fully developed homogeneous turbulent flow at high Reynolds numbers (Re). Assuming the aggregate size to be smaller than the Kolmogorov length-scale $\eta$, the hydrodynamic stress is associated with the *local* energy dissipation rate $\varepsilon$, whose fluctuations are highly intermittent. It has been recently shown in [27] that the breakup kinetics in this regime is governed by the frequency at which $\varepsilon$ exceeds a critical value ($\varepsilon_{crit}$) corresponding to the critical stress for aggregate breakup. Further, it was found that the obtained long-time or steady-state mass scaling with shear rate from solving the full fragmentation equation (with an appropriate breakup rate kernel) is equivalent to the scaling $\lim_{t \to \infty} R_g(t) \sim \dot{\gamma}^p \sim \varepsilon_{crit}^{p/2} \sim <\varepsilon>^{p/2} [\eta/l_0^{4(\alpha_{min}-1)/(\alpha_{min}+3)}]^{p/2}$, derived in [28], which makes use of the multifractal description of turbulence and where $\dot{\gamma}$ is the turbulent shear rate. (Note the difference of a factor two with the definition of $p$ given in Ref. [27]). The $\eta/l_0$ is the length-scale separation, $\alpha_{min}$ is the lower limit of the scaling exponent of the multifractal spectrum (corresponding to the harshest turbulent event), while $p$ is treated as a lumped (free) parameter accounting for the mechanical response of the aggregate [27,28]. This relation is equivalent to replacing $\sigma \sim \dot{\gamma} \sim \varepsilon_{crit}^{1/2}$ in Eqs. (7). It has been assumed that after a sufficiently long time all aggregates have sampled the whole multifractal spectrum, including the highest local velocity gradients (associated with $\varepsilon_{crit}$ and $\alpha_{min}$) which determine the critical stress. Thus, on the basis of these considerations and of Eq. (8) as derived here, the lumped power-law exponent of Ref. [27] can be identified as $p = -2/(d-d_f)[2(\beta+1)-\nu]$ (with the aforementioned difference of a factor two in the definition of $p$ with respect to [27]). Due to the large scale heterogeneity in stirring devices, near the impeller blades the local energy dissipation rate, $\varepsilon$, can reach values orders of magnitude higher than the volume averaged value, $\langle\varepsilon\rangle$ [23,31]. We note that $\langle\varepsilon\rangle$ represents a flow-intensity parameter easy accessible e.g. by torque measurements. Taking this into consideration, also the length scale separation $\eta/l_0$ can be estimated by means of CFD simulations, as a function of $\langle\varepsilon\rangle$. In stirring devices, the length-scale separation is a power-law of $\langle\varepsilon\rangle$, $\eta/l_0 \sim \langle\varepsilon\rangle^\theta$. Hence, we derive

$$\lim_{t\to\infty} X(t) \equiv X^{(s)} \sim \langle\varepsilon\rangle^{\frac{-d_f}{(d-d_f)[2(\beta+1)-\nu]}\left[1+\frac{4\theta(\alpha_{\min}-1)}{\alpha_{\min}+3}\right]} \qquad (8)\text{-a}$$

$$\lim_{t\to\infty} R_g(t) \equiv R_g^{(s)} \sim \langle\varepsilon\rangle^{\frac{-1}{(d-d_f)[2(\beta+1)-\nu]}\left[1+\frac{4\theta(\alpha_{\min}-1)}{\alpha_{\min}+3}\right]} \qquad (8)\text{-b}$$

We now compare predictions of Eqs. (8) with experimental data from [23], referring to aggregate breakup under turbulent flow in a stirring apparatus ($1.2\times10^4 < \text{Re} < 6\times10^4$), where our assumptions are fulfilled. Again, dilution was such that aggregate breakup is unaffected by aggregation phenomena. The colloid system and particle radius was the same as in the experiments reported here, i.e. $a \simeq 405\,\text{nm}$. The relation $\eta/l_0 \sim \langle\varepsilon\rangle^{-0.234}$ was determined from numerical simulations of the flow field in the stirring apparatus of [23]. Also in this case, the aggregate fractal dimension does not change appreciably with the average flow intensity, thus with $\langle\varepsilon\rangle$, and was estimated in [23] as $d_f = 2.62\pm0.2$. The value $\alpha_{\min} \simeq 0.12$, a property of the multifractal spectrum corresponding to the most intense turbulent event, has been employed according to [27]. Thus, the comparison between predictions of Eq. (8)-a, with $d_f = 2.63$, and the experimental data of [23] is shown in Fig. 3(a). The agreement between the predicted scaling and the experimental trend, once more, is excellent. We note that the effect of intermittency, as already pointed out in [27], amounts to increasing the absolute value of the power-law exponent by only 24% of the value for laminar flow (the latter given by Eq. (6)). This is a rather small correction. The uncertainty in the experimentally measured fractal dimension in [23], requires an independent estimate. By the same analysis as explained above, we found that in order to obtain the best quantitative agreement between the $I(0)$ values measured experimentally (in [23]) and those calculated from computer-generated aggregates, as shown in Fig. 3(b), the fractal dimension must be $d_f = 2.63$, thus consistent with the value $d_f = 2.62 \pm 0.18$ experimentally found in [23]. This justifies the $d_f$ value used for the comparison with experiments, and confirms the general good agreement between our scaling approach and the experimental data also in the case of turbulent flows. Also in this case, as shown in Fig. 3(b), the error that one makes if multiple scattering is not taken into account is very large.

### C. Breakup exponent versus fractal dimension

Our findings and the emerging picture are summarized in Fig. (4), as a breakup exponent versus fractal dimension diagram, where our model predictions are compared with experimental and simulation data from several authors. We observe that upon increasing $d_f$ in the range $2.4 < d_f < 2.8$, the breakup exponent $p$ decreases very slowly from about -0.3 to about -0.8. Further, using $v = -0.4$ rather than $v = -1/(d - d_f)$ does not lead to significant differences in this regime. However, starting from $d_f \simeq 2.8$ and getting closer to the homogeneous limit $d_f \to d$, the curves for the two values of $v$ differ substantially. In particular, we expect the scaling with $v = -0.4$ to be the more realistic one in this regime as it recovers the correct limit $p \to -\infty$ at $d_f = d$, where the stress must eventually become independent of the aggregate size. Thus, in the regime $2.4 < d_f < 2.8$, which is still dominated by fractality, hence by a significant decay of the inner density with the linear size of the aggregate, our model predictions are in excellent agreement not only with the experimental data from our lab, as shown above, but also with the simulations of Higashitani et al. [29]. In this regime, if the fractal dimension does not change upon breaking up (as observed experimentally), the fragments generated upon increasing the hydrodynamic stress are significantly denser than the precursor aggregate (since they are smaller), thus they better withstand the hydrodynamic stresses. This leads to a considerable mitigation of the breakup-induced decrease of the average stable size upon increasing the stress which is reflected in low absolute values (<1) of the breakup exponent $p$. On the other hand, in the limit of weakly fractal or quasi-homogeneous and eventually homogeneous (non-fractal) aggregates ($d_f > 2.8$), once the critical stress is applied, any fragment will undergo breakup regardless its size, thus resulting in a value of critical stress practically independent of the aggregate size, hence in high ($>>1$) absolute values of $p$. This limit is captured equally well by our model if the scaling with $v = -0.4$, valid indeed for non-fractal disordered solids, is used, as confirmed by the agreement between the our model predictions and the experimental data of Refs. [13] and [30] shown in Fig. (4). When $d_f \simeq d$, the fracture criterion of non-fractal aggregates should be more properly given in terms of the volume fraction as $\sigma \sim \phi^{(\beta+1)-v/2}$. We also note that this picture remains valid in turbulent flows, at least for the fully-developed fractal regime in which experimental data are available (inset of Fig. (4)).

## VI. CONCLUSION

In sum, we have shown the evidence that the breakup of dense colloidal aggregates with $d_f \simeq 2.4-3.0$, under an applied laminar or turbulent flow, can be understood in terms of a critical hydrodynamic stress associated with a critical strain energy given by the bond-energy required for unstable crack-propagation in the aggregate. The inner dense amorphous structure of such aggregates as those formed under flow conditions is responsible for a brittle mechanical response typical of glassy materials. However, with aggregates in the fractal dimension range $d_f \simeq 2.4-2.8$, owing to the significant decay of volume fraction with the linear size of the aggregate, the observed decrease of the stable size with the hydrodynamic stress is made much less steeper in comparison with homogenous (non-fractal) solids for which the critical stress is practically independent of the aggregate size. This picture has been found to agree well with experimental results from our lab as well as with simulations and experiments from the literature, in both laminar and turbulent flows. These findings will be used in future work to improve our current understanding of the microscopic origin of the peculiar rheological properties of strongly-sheared interacting colloids where breakup greatly affects the structure-formation and structure-failure processes (by limiting the former and enhancing the latter), whose interplay is responsible for puzzling behaviours such as thixotropy and rheopexy [4].

ACKNOWLEDGMENTS

Lyonel Ehrl is gratefully acknowledged for the tuneable fractal dimension code. Financial support from Swiss National Foundation (grant. No. 200020-113805/1) is gratefully acknowledged. A. Z. thanks Dr. E. Del Gado for many fruitful discussions.

FIGURE 1

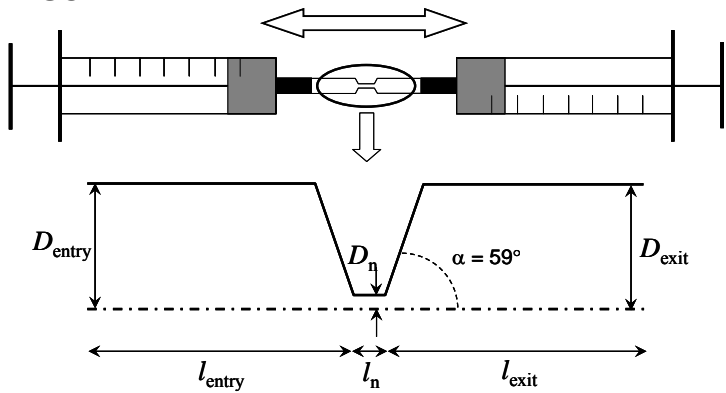

FIG. 1. Geometry of the convergent-divergent multi-pass channel used to realise the extensional flow to study aggregate breakup.

FIGURE 2

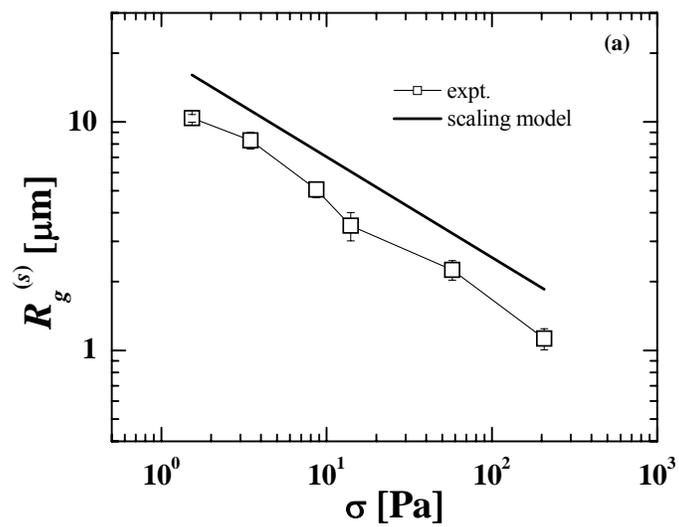

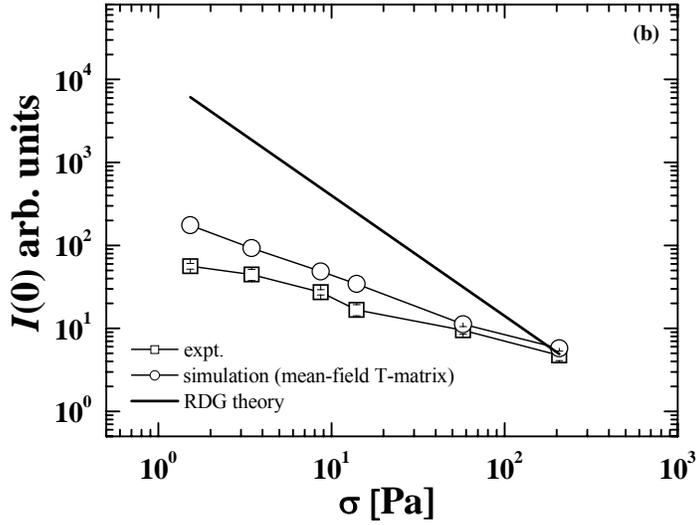

FIG. 2. (a) Comparison between experimental data of steady-state aggregate size in extensional flow (see Text) and the scaling prediction from Eq. (7)-a with $d_f = 2.7$. (b) Comparison between the steady-state zero-angle scattered light intensity, $I(0) \sim X^{(s)}$, measured experimentally by SALS (squares), and simulations of $I(0)$ for computer-generated aggregates with $d_f = 2.7$ (circles). Also shown is the scaling, as from Eq. (7)-b, without accounting for multiple scattering.

FIGURE 3

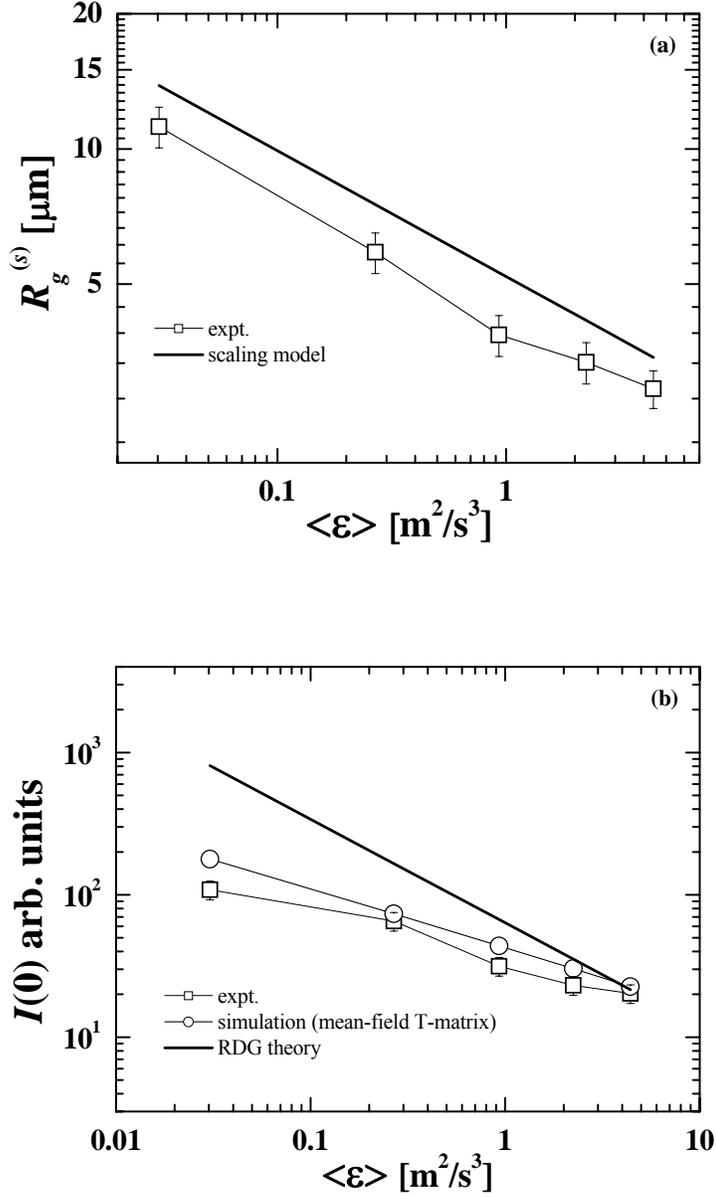

FIG. 3. (a) Comparison between experimental data of steady-state aggregate size in turbulence from Ref. [23] and the scaling prediction from Eq. (8)-a with $d_f = 2.63$ (solid line). (b) Comparison between the zero-angle scattered light intensity, $I(0) \sim X^{(s)}$, measured experimentally by SALS in Ref. [23] (squares), and simulations of $I(0)$ for computer-generated aggregates with $d_f = 2.63$ (circles). Also shown is the scaling without accounting for multiple scattering, as from Eq. (8)-b.

FIGURE 4

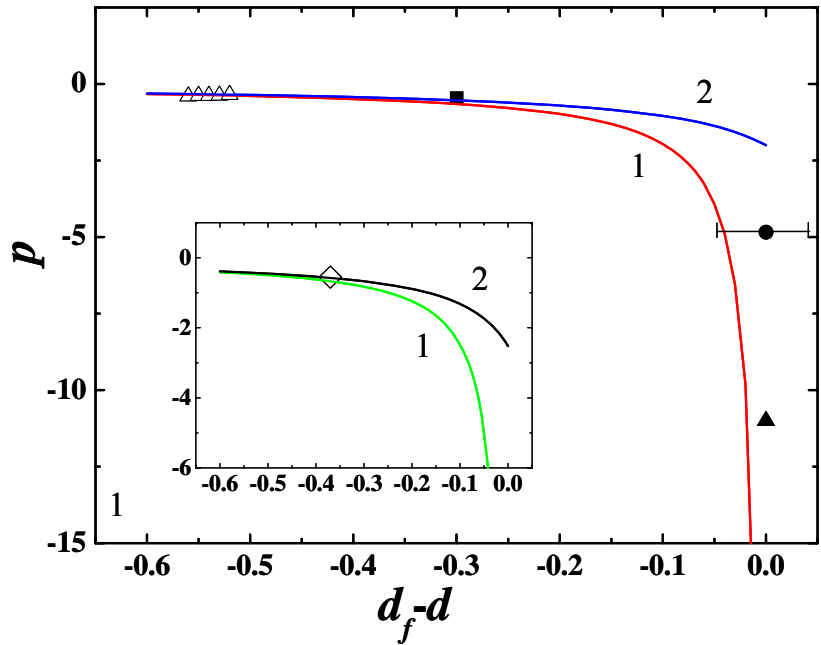

FIG. 4. (Color online) Map of the breakup exponent $p$ as a function of fractal dimension for laminar flows. Curve 1 and 2 are given by $p = -2/(d-d_f)[2(\beta+1)-\nu]$, with $\nu = -0.4$ and $\nu = -1/(d-d_f)$ respectively. Symbols: (Δ) simulation data from Ref. [29]; (■) experimental data from the present work; (●) experimental data from Ref. [30]; (▲) experimental data from Ref. [13]. Inset: same comparison for turbulent flow. Symbols: (◊) experimental data from Ref. [23].